\documentclass[aps,prb,twocolumn,superscriptaddress]{revtex4}

\usepackage{graphicx}

\begin{document}


\title{Local Measurements of Magnetization in $\mathrm{Mn_{12}}$
Crystals}


\author{Nurit Avraham}
\affiliation{Dept. of Condensed Matter Physics, The Weizmann
Institute of Science, Rehovot 76100, Israel}
\affiliation{Physics
Department, City College of the City University of New York, New
York, NY 10031}
\author{Ady Stern}
\affiliation{Dept. of Condensed Matter Physics, The Weizmann
Institute of Science, Rehovot 76100, Israel}
\author{Yoko Suzuki}
\author{K. M. Mertes}
\author{M. P. Sarachik}
\affiliation{Physics Department, City College of the City
University of New York, New York, NY 10031}
\author{E. Zeldov}
\author{Y. Myasoedov}
\author{H. Shtrikman}
\affiliation{Dept. of Condensed Matter Physics, The Weizmann
Institute of Science, Rehovot 76100, Israel}
\author{E. M.  Rumberger}
\author{D. N. Hendrickson}
\affiliation{Dept. of Chemistry and Biochemistry, University of
California at San Diego, La Jolla, CA 92093}
\author{N. E. Chakov}
\author{G. Christou}
\affiliation{Department of Chemistry, University of Florida,
Gainesville, FL 32611 }

\date{\today}

\begin{abstract}
The spatial profile of the magnetization of $\mathrm{Mn_{12}}$
crystals in a swept magnetic field applied along the easy axis is
determined from measurements of the local magnetic induction along the
sample surface using an array of Hall sensors. We find that the
magnetization is not uniform inside the sample, but rather shows some
spatial oscillations which become more prominent around the resonance
field values. Moreover, it appears that different regions of the sample
are at resonance at different values of the applied field and that the
sweep rate of the internal magnetic induction is spatially non-uniform. We
present a model which describes the evolution of the
non-uniformities as a function of the applied field. Finally we
show that the degree of non-uniformity can be manipulated by
sweeping the magnetic field back and forth through part of the
resonance.
\end{abstract}


\maketitle

Molecular magnets, or single molecule magnets, are typically
composed of magnetic cores surrounded by organic complexes.  With
a total spin of  $S=10$ and strong easy axis anisotropy
\cite{caneschi91,sessoli93a,sessoli93b,sessoli95,novak95a,paulsen95a,novak95b,paulsen95b},
$[\mathrm{Mn_{12}O_{12}(CH_{3}COO)_{16}(H_{2}O)_{4}]\cdot2CH_{3}COOH\cdot4H_{2}O}$
(generally referred to as $\mathrm{Mn_{12}}$-acetate) has received
a great deal of attention.  When crystallized, $\mathrm{Mn_{12}}$
forms a tetragonal lattice \cite{lis} with the easy magnetization
direction along the c-axis. The magnetic cores are well separated
\cite{lis}, so that exchange interactions between the molecules
are negligible \cite {novak95a,paulsen95a,novak95b,paulsen95c}.
The main terms in the spin Hamiltonian of $\mathrm{Mn_{12}}$ are
given by
\begin{eqnarray}\label{Hamiltonian}
 H=-DS_{z}^{2}-g_{z}\mu_{_{B}}B_{z}S_{z}-AS_{z}^{4}\nonumber
\end{eqnarray}
The first term is the magnetic anisotropy energy, with $D =
0.548(3)\mathrm{K}$ while the third term represents the next
higher-order term in longitudinal anisotropy, with  $A =
1.173(4)\times10^{-3} \mathrm{K}$ \cite{barra,hill,mirebeau}.  The
second term is the Zeeman coupling to a magnetic field $B_{z}$
along the easy-axis. In the absence of a magnetic field, this
Hamiltonian can be represented as a symmetric double well
potential with a set of energy levels corresponding to the
$(2S+1)=21$ allowed values of the quantum number $S_{z}$.  A
magnetic field $B_{z}$ lifts the degeneracy of $\pm S_{z}$ states
on opposite sides of the potential barrier, and a strong field
$B_{z}$ causes most of the molecules to occupy states in one of
the wells. When $B_{z}$ is swept, e.g., from positive to negative
values, molecules relax from one side of the potential barrier to
the other either via thermal activation over the barrier or by
quantum tunneling through the barrier. Measurements below the
blocking temperature of 3K show resonant tunneling of the
magnetization, manifested as a series of steep steps in the
hysteresis loop at roughly equal intervals of magnetic field \cite
{friedmanJAP,friedmanPRL,hernandez,thomas}. The steps occur at
magnetic field values where states on opposite sides of the
barrier have the same energy, and are at ``resonance''.

One of the common experimental techniques to study the tunneling
of the magnetization is to measure the magnetic response of the
crystal when an external magnetic field $H_{a}$ is applied
parallel to the easy axis, and is swept through a series of
resonances along the hysteresis loop. To date, studies of extended
samples have provided the magnetic induction (and its variation in
a swept magnetic field) as averages over the whole sample, or they
have assumed that values measured in a limited region of space
represent the average. As such, these measurements do not provide
any information on the spatial profile of the magnetization in the
sample, or on how the relaxation process propagates spatially
within the sample.

The experiment reported here utilizes an array of sensors to
measure the spatial variation of the magnetization in a
$\mathrm{Mn_{12}}$ crystal as $H_{a}$ is swept. Within our model
we find the magnetization to be significantly non-uniform along
the sample, with the non-uniformity being enhanced within the
steps. Furthermore, it appears that different regions of the
sample are at resonance at different values of the applied field,
and that the sweep rate of the internal magnetic induction is
spatially non-uniform. We demonstrate that the degree of
non-uniformity can be manipulated by sweeping the magnetic field
back and forth through part of the resonance.

\begin{figure}
\includegraphics[width=0.4\textwidth]{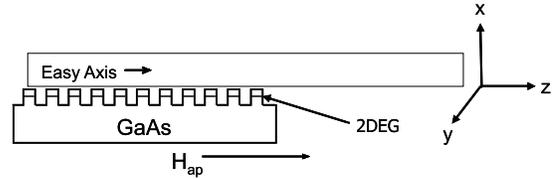}
\caption{\label{fig1}Schematic diagram of the experimental setup.
The $\mathrm{Mn_{12}}$ crystal, with a typical size of $z=400$
$\mu m$, $y=100$ $\mu m$ and $x=40$ $\mu m$ is placed directly on
the surface of an array of $10\times10$ $\mu m^{2}$ sensors that
are $10$ $\mu m$ apart. The sensors cover only half the sample and
measure the perpendicular component ($B_{x}$) of the magnetic
induction at the surface.}
\end{figure}

The local induction of single $\mathrm{Mn_{12}}$ crystals of
typical size $400\times100\times40$ $\mu m ^{3}$ was measured
using an array of eleven Hall sensors \cite{hallbars}. The active
layer of these sensors is a two dimensional electron gas formed at
the interface of GaAs/AlGaAs heterostructure. The samples were
mounted onto the surface of an array of $10\times10$ $\mu m ^{2}$
sensors that were $10$ $\mu m$ apart, with the easy-axis parallel
to the \emph{z}-direction and to the applied field $H_{a}$, as
shown in Fig. 1. The eleven sensors measure $B_{x}$, the
\emph{x}-component of the magnetic induction due to the
magnetization of the crystal. Since the two dimensional electron
gas resides only $0.1$ $\mu m$ below the surface, the induction
measured by the sensors is practically equal to the induction at
the crystal surface. As shown in Fig. 1, the center of the crystal
was aligned with the array so that ten sensors probed the field
along half the crystal length and the last sensor measured the
field at the edge of the sample.

Fig. 2a and 2b show two local hysteresis loops of $B_{x}(H_{a})$,
at $T = 0.3 K$, measured simultaneously at the edge of the sample
and close to its center. The curves were obtained starting from a
fully magnetized state by sweeping $H_{a}$ from $-6$T up to 6T and
back. The loop measured locally at the edge of the sample (Fig.
2a) resembles previously measured magnetization loops. When
increasing $H_{a}$ from zero to 6T, $B_{x}$ increases
monotonically toward its positive saturation value while
displaying a series of steep steps at roughly equal intervals of
magnetic field, due to resonant tunneling of the magnetization.

In contrast, the magnetic induction measured close to the center
of the sample (shown in Fig. 2b) displays non-monotonic behavior as a
function of $H_{a}$. The resonances are not manifested by a series
of steps, but rather by a series of peaks and dips whose width is
larger than that of the steps in Fig. 2a. Similar behavior is
observed at other sensors located between the edge and the center
of the sample. In general, $B_{x}$ is largest close to the edge
and decreases towards the center of the sample. The non-monotonic
dependence on $H_{a}$, however, is most pronounced close to the
center and gradually diminishes when proceeding
towards the edge of the sample.
\begin{figure}
\includegraphics[width=0.35\textwidth]{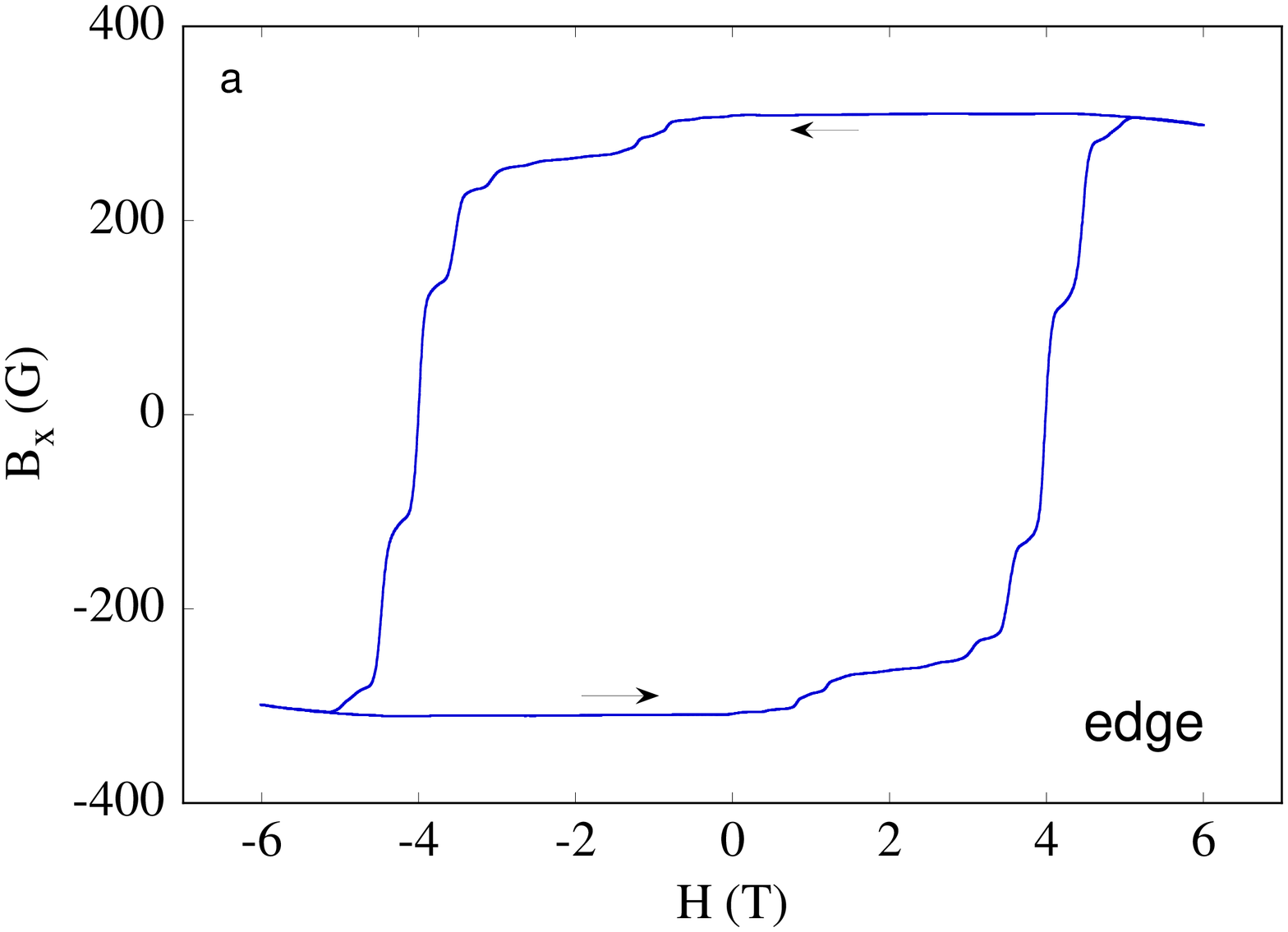}\\
\includegraphics[width=0.35\textwidth]{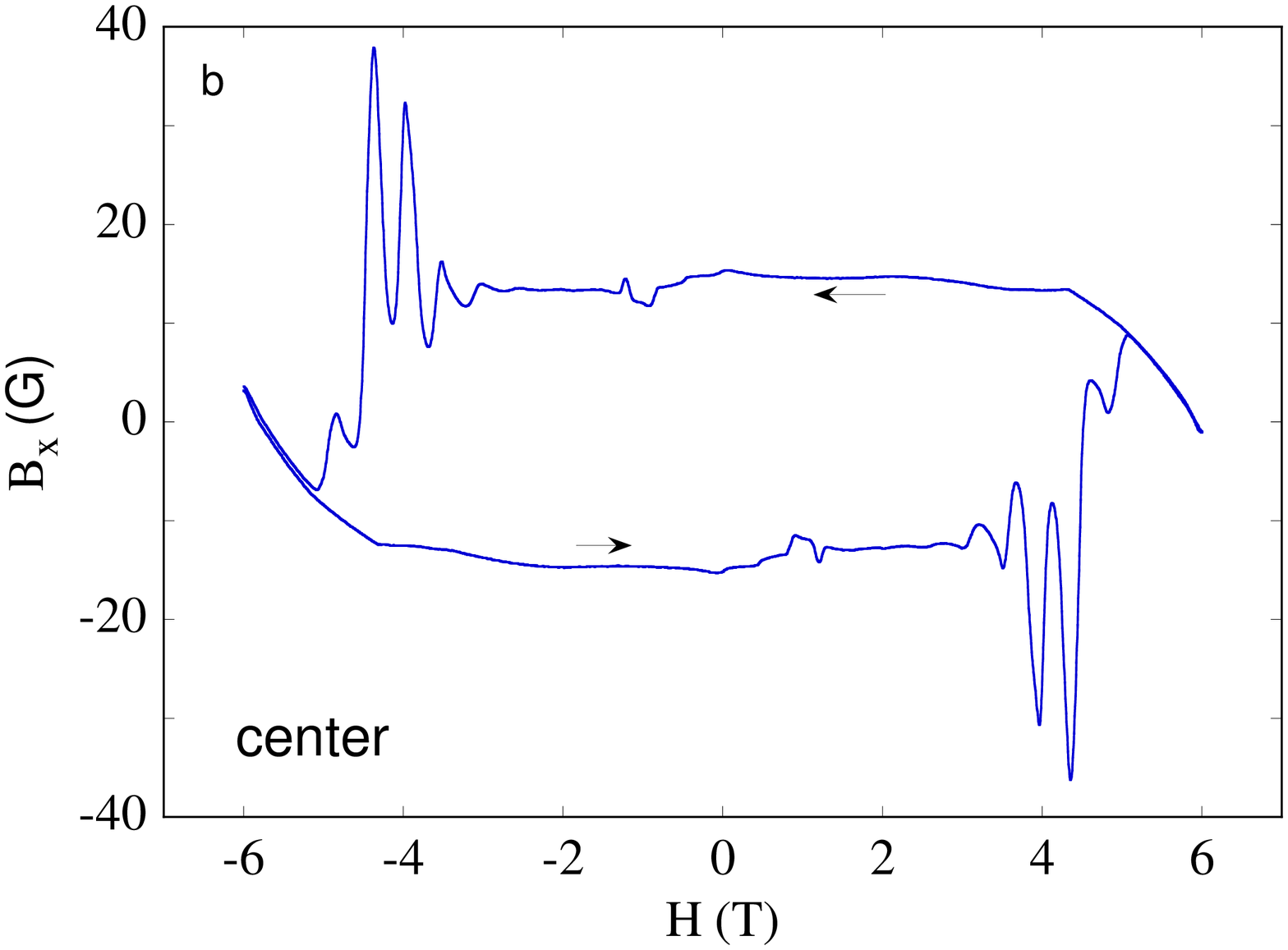}
\caption{\label{fig2} Local hysteresis loops of $B_{x}$ as a
function of $H_{a}$ ($T = 0.3\mathrm{K}$) , measured
simultaneously at two different positions on the sample: (a) close
to the edge and, (b) close to the center of the sample.}
\end{figure}
The remarkable difference in the dependence of $B_{x}$ on $H_{a}$
observed by the different sensors contradict the assumption that
the magnetization is uniform inside the sample. In general
\cite{jackson},
$B_{x}(\vec{r},H_{a})=\int\vec{dr'}M(\vec{r'},H_{a})f{(\vec{r}-\vec{r'})}$,
where $M(\vec{r'},H_{a})$ is the magnetization at point
$\vec{r'}$, that is assumed to be parallel to the easy axis, and
\begin{eqnarray}\label{f}
  f(\vec{r}-\vec{r'})=-\frac{3(z-z')(x-x')}{{|r-r'|^{5}}}\nonumber
\end{eqnarray}
If $M=M_{0}(H_{a})$ is independent of $\vec{r'}$, we get
$B_{x}(\vec{r},H_{a})=M_{0}(H_{a})\int\vec{dr'}f(\vec{r}-\vec{r'})$,
so that the $B_{x}(H_{a})$ curves measured by all sensors should
collapse onto a single curve when multiplied by a scaling factor.
As the inner curves are non-monotonic while the edge curves are
monotonic, this scaling cannot be realized, and the magnetization
is therefore non-uniform.

The non-uniformity of the magnetization is the source of the
non-monotonicity displayed in Fig. 2b. Spins pointing in the same
direction that are located on opposite sides of a sensor make
\emph{opposite} contributions to $B_{x}$. Due to these opposite
contributions, the magnetic induction $B_{x}$ measured by a sensor
located close to the center of the sample is very small when all
the spins in its surrounding point in the same direction, but is
enhanced when the magnetization on the two sides is different.
This enhancement may bring $B_{x}$ to a value larger than its
value when the sample becomes fully magnetized, as shown in Fig.
2b. Now, if the relaxation rates on the two sides of the sensor
are different, the non-uniformity in the magnetization varies, and
so does $B_{x}$. The direction in which $B_{x}$ varies with
$H_{a}$ indicates the relative magnitudes of the relaxation rates
on the two sides of the sensor. The situation is different for a
sensor located near the edge of the sample, where all spins are on
one side, and therefore their contributions to $B_{x}$ add. For
this sensor, inhomogeneities in the magnetization only broaden the
steps, but do not result in non-monotonic behavior. This
qualitative picture is further discussed below in terms of an
equation for the time evolution of the local magnetization.

We now proceed to extract the spatial profile of the magnetization
from the profile $B_{x}(z)$ measured by the sensors. Assuming that
the sample has a perfect rectangular shape, and the magnetization
is uniform along the \emph{x} and \emph{y} directions, the
induction $B_{x}$ can be written as
\begin{equation}\label{eq.1}
  B_{x}(z)=-\int dz'F(z-z')\frac{\partial M(z')}{\partial z}
\end{equation} where
\begin{eqnarray}\label{eq.2}
  F(z-z')=\log\frac{w_{1}+\sqrt{w_{1}^{2}+d_{1}^{2}+(z-z')^{2}}}
  {w_{1}+\sqrt{w_{1}^{2}+d_{2}^{2}+(z-z')^{2}}}-\nonumber \\\log\frac{w_{2}+\sqrt{w_{2}^{2}+d_{1}^{2}+(z-z')^{2}}}{w_{2}+\sqrt{w_{2}^{2}+d_{2}^{2}+(z-z')^{2}}}
\end{eqnarray}
For the samples we used, $d_{1} = 0.1$ $\mu m$ is the distance in
the \emph{x} direction between the sample surface and the 2DEG,
$d_{2}= 40$ $\mu m$ is the distance between the 2DEG and the
sample's other surface, and $w_{1} = -5$ $\mu m $, $w_{2} = 85$
$\mu m$ are the distances between the center of the sensors and
the edges of the sample in the \emph{y}-direction.

$F(z-z')$ is a short-range function, whose characteristic decay
length is typically $10$ $\mu m $ for our geometry. Eq.
(\ref{eq.1}) implies that the magnetic field $B_{x}$ measured by a
sensor on the sample surface is proportional to the derivative
$\frac{\partial M}{\partial z}$ in the vicinity of the sensor, and
therefore Figs. 2a and 2b approximate the dependence of
$\frac{\partial M}{\partial z}$ on $H_{a}$ at the edge of the
sample and close to the center of the sample, respectively.

Assuming that the magnetization varies smoothly over the scale of
our sensors, we interpolate the measured $B_{x}(z)$ and
numerically invert the kernel $F(z-z')$ to extract the
magnetization profile $M(z,H_{a})$. Figure 3 shows the resulting
magnetization as a function of position for various values of
$H_{a}$. Since our sensors cover only half the sample, we rely on
the (nearly) symmetric shape of the samples and assume the
magnetization to be symmetric with respect to the sample center.
The non-uniformity of the magnetization and its evolution as a
function of $H_{a}$ are clearly demonstrated. Focusing first on
the central part of the sample, we note that at $-5.7$T, where the
sample is presumably fully magnetized, the magnetization is
approximately uniform. When increasing the field to 3.6T the
magnetization starts to develop a small amount of non-uniformity,
manifested by the small variations of the magnetization from the
average value. These variations continue to develop and become
more pronounced at 3.9T, when the magnetization in most of the
sample shows a high relaxation rate. The non-uniformities then
decay somewhat at 4.14T, where the sample is out of resonance, and
become prominent once more at the next resonance, at 4.3T. When
increasing the field further to 5.7T the non-uniformities
disappear and the magnetization becomes uniform again.
Non-uniformities similar to those presented in Fig. 3 were
observed in several samples, where they are enhanced when most of
the sample is at resonance, and become minor away from resonances.
\begin{figure}
\includegraphics[width=0.4\textwidth]{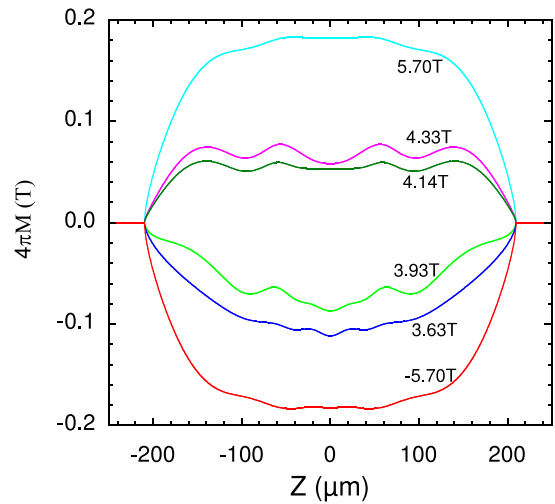}
\caption{\label{fig3} The calculated profile of the magnetization
plotted as a function of the distance from the center of the
sample, at six different values of the applied field. Note the
enhanced non-uniformity at fields close to resonances, at 3.9T and
4.33T.}
\end{figure}

At the very edge of the sample the magnetization is expected to
change abruptly from zero outside the sample to a non-zero value
inside the sample. We believe that the gradual variation observed
at all field values results from two major sources. First, the
field $B_{x}$ at the edge of the sample varies significantly over
the length of one sensor, and therefore the average field measured
by the edge sensor may deviate appreciably from its local value.
Second, the calculation of the magnetization is carried out
assuming the sample has a perfect rectangular shape. Deviations
from this shape, for example edges that are not perpendicular to
the plane of the sensors, could also contribute to the rounding of
the data at the edges. Note, however, that these effects cannot
introduce the observed non-uniformities in the center of the
sample. We also note that the overall scale of the magnetization
obtained from our calculation is not very different from that
expected: at full magnetization, $4\pi M$ is 0.12T for
$\mathrm{Mn_{12}}$ crystals, while the maximum value shown in Fig.
3 is 0.18T.

The non-uniformity of the magnetization implies that  $B_{z}$, the
internal magnetic field in the z direction, is spatially non-uniform.
Consequently, different parts of the sample enter the resonance at
different values of $H_{a}$. We will now demonstrate that the internal
sweep rate $\frac{\partial B_z}{\partial t}$ is spatially non-uniform as
well, so that different parts of the sample spend different times at
resonance.
Under the assumption used to extract $M(z)$, $B_{z}$ inside the
sample can be written as
\begin{equation}\label{eq.3}
  B_{z}(z)=H_{a}+4\pi M(z)+\int dr'\frac{\partial M(r')}{\partial z'}\frac{z-z'}{|r-r'|^{3}}
\end{equation}
Since all the terms and their time dependence are available from
our analysis, we can directly derive the local sweep rate
$\frac{\partial B_{z}}{\partial t}$.

This is shown in Fig. 4 as a function of position in the central
part of the sample for different values of $H_{a}$. Two major
observations are of particular interest. The first is that at 3.9T
and at 4.3T, when the sample is at resonance, the internal sweep
rate is much higher than the nominal sweep rate of the external
field. For example, $\frac{\partial B_z}{\partial t}$ is about 2.5
times larger than $\frac{\partial H_{a}}{\partial t}$ at 3.9T. In
contrast, at 3.6T and 4.1T, when the sample is out of resonance,
the sweep rates of $B_{z}$ and $H_{a}$ are approximately equal.
The second major observation is that at resonance fields,
$\frac{\partial B_z}{\partial t}$ is highly non-uniform and can
vary by as much as $20\%$ over a distance of the order of $50$
$\mu m$. This means not only that different regions enter the
resonance at different $H_{a}$, but also that the times spent by
local regions at resonance can be significantly different.

\begin{figure}
\includegraphics[width=0.4\textwidth]{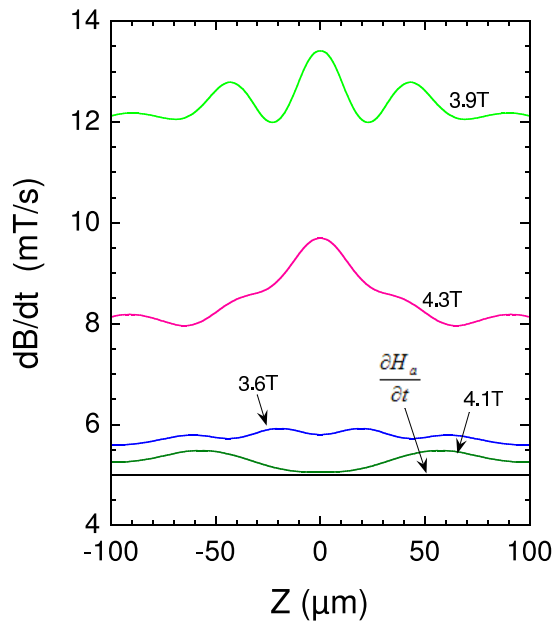}
\caption{\label{fig4} Calculated sweep rate of the internal field,
$\frac{\partial B_{z}}{\partial t}$, plotted as a function of the
distance from the center of the sample, for the same field values
as in Fig. 3.  Note that when most of the sample is at resonance,
at 3.9T and 4.3T, the internal sweep rate $\frac{\partial
B_z}{\partial t}$ is much higher than the external sweep rate,
$\frac{\partial H_{a}}{\partial t}$. Moreover, at these fields
$\frac{\partial B_z}{\partial t}$ is significantly non-uniform
inside the sample. When the sample is out of resonance, at 3.6T
and 4.1T, the internal sweep rate is approximately uniform inside
the sample and is very close to the sweep rate of the external
field.}
\end{figure}

As we show below, the observed spatial non-uniformities and their
evolution with $H_{a}$ can be explained by a simple model
describing the relaxation of the magnetization along the
hysteresis loop. Initially, at 6T, the magnetization is uniform
inside the sample. However the relaxation rate is non-uniform due
to the spatial variation of the internal magnetic field and due to
molecular and structural disorder. Therefore, as we sweep the
magnetic field to $-6$T (in the opposite direction) non-uniformity
in the magnetization begins to develop. Under certain conditions,
and particularly within resonances, the non-uniformity in the
magnetization grows with time, despite the decay in the
magnetization itself. The magnetization is non-uniform even when
the sample is out of resonance and the relaxation rates are
negligible. We attribute the non-uniformity in the magnetization
off resonance to the spatial variation of the sweep rate (Fig. 4)
that causes different regions of the sample to spend different
times at resonance.

In our model the relaxation of the space- and time-dependent
magnetization $M(z,t)$ is governed by the equation
\begin{equation}\label{eq.4}
  \frac{\partial M}{\partial t}=-(M-M_{0})\Gamma
\end{equation}
where $M_{0}$ is the equilibrium magnetization and $\Gamma$ is the
relaxation rate, which near a resonance has a general form
\begin{equation}\label{eq.5}
\Gamma=\Gamma_{0}\frac{\alpha^{2}}{(B_{z}-B_{r})^{2}+\alpha^{2}}
\end{equation}
where $B_{r}$ is the magnetic induction at the center of the
resonance, and $2\alpha$ is the resonance width. Molecular
disorder introduces variations of $\Gamma_{0}$, $B_{r}$ and
$\alpha$ between different molecules. However, since the
distribution of these parameters does not depend on position, we
use their average values. The non-uniformity of the magnetization
is characterized by its spatial derivative $\frac{\partial
M}{\partial z}$. We may use Eq. (\ref{eq.4}) to analyze the time
evolution of $\frac{\partial M}{\partial z}$. To that end, we take
the derivative of (4) with respect to \emph{z}, and approximate
$B(z)\approx H_{a}+4\pi M(z)$(i.e., we neglect the third term in
Eq. (\ref{eq.3}), which is small in the central part of the
sample). We obtain
\begin{equation}\label{eq.6}
  \frac{\partial}{\partial t}\frac{\partial M}{\partial
  z}=-[\Gamma+4\pi(M-M_{0})\frac{\partial \Gamma}{\partial B}]\frac{\partial M}{\partial z}
\end{equation}
Although the dependence of $\Gamma$ on $M(z)$ makes this equation
a non-linear differential equation, the conditions under which
non-uniformities in the magnetization grow with time are rather
transparent. The question of whether the magnitude of
$\frac{\partial M}{\partial z}$ increases or decreases with time
is determined by the balance between the two terms on the right
hand side of this equation. The first term acts to suppress the
magnitude of $\frac{\partial M}{\partial z}$, i.e., the magnitude
of the non-uniformity. The effect of the second term, on the other
hand, depends on the sign of $(M-M_{0})\frac{\partial
\Gamma}{\partial B}$. When this sign is negative, this term tends
to enhance the magnitude of the non-uniformity. If strong enough,
it may overcome the effect of the first term. When that happens,
the magnetization $M(z)$ relaxes towards $M_{0}$, but  its
non-uniformity $|\frac{\partial M}{\partial z}|$ increases with
time.

From Eq. (\ref{eq.6}), it is easy to verify that as long as
$|M-M_{0}|<\frac{\alpha}{4\pi}$,
\begin{equation}\label{eq.7}
  \Gamma+4\pi(M-M_{0})\frac{\partial \Gamma}{\partial B}>0
\end{equation} for any $B_{z}$
and hence the non-uniformity is suppressed with time. However,
when $|M-M_{0}|>\frac{\alpha}{4\pi}$, there is a range of $B_{z}$
\begin{eqnarray}\label{eq.8}
  (M-M_{0})-\sqrt{(M-M_{0})^{2}-\frac{\alpha^{2}}{(4\pi)^{2}}}<\frac{(B_{z}-B_{r})}{4\pi}\linebreak
  <\nonumber\\(M-M_{0})+\sqrt{(M-M_{0})^{2}-\frac{\alpha^{2}}{(4\pi)^{2}}}
\end{eqnarray}
at which $\Gamma+4\pi(M-M_{0})\frac{\partial \Gamma}{\partial
B}<0$ and the non-uniformity is enhanced with time. We consider,
for concreteness, the case where $H_{a}$ is swept down from 6T to
$-6$T through one of the resonances. In this case $M_{0}$ is
negative, and $M-M_{0}$ is positive. If
$M-M_{0}>>\frac{\alpha}{4\pi}$, the range (\ref{eq.8}) occupies
the first half of the resonance, where $B_{z}>B_{r}$ . As $H_{a}$
is swept down, the non-uniformity grows during the first half of
the resonance while $B_{z}>B_{r}$, and then is suppressed again
when $B_{z}$ drops below $B_{r}$. Such behavior is indeed observed
in Fig. 3. In general, non-uniformities are always enhanced on
approaching resonance, reaching maximum close to resonance and
then suppressed after the resonance.   Our measurements show
resonances whose half-widths are about 0.045T. However, the
measured half-widths are larger than the intrinsic half-width
$\alpha$, due to the effects of molecular disorder and hyperfine
fields. The non-uniformities are found experimentally to evolve at
the resonances where $4\pi(M-M_{0})>0.06$T, consistent with the
model above.

Eq. (\ref{eq.5}) and the discussion following it suggest that the
non-uniformity in the magnetization may be enhanced if the applied
field $H_{a}$ is tuned to a value where
$\Gamma+4\pi(M-M_{0})\frac{\partial \Gamma}{\partial B}$ is
negative at least in some parts of the sample, and is kept there
for some time \cite{wernsdorfer}. Our measurements show that this
is indeed the case. To identify the proper value of $H_{a}$, we
note again that $B_{x}$ is proportional to $\frac{\partial
M}{\partial z}$ and therefore the observed slope $\frac{\partial
B_{x}}{\partial H_{a}}$ measures the local slope
$\frac{\partial}{\partial z}\frac{\partial M}{\partial t}$. Thus,
when $H_{a}$ is swept back and forth over a region where both
$\frac{\partial B_{x}}{\partial H_{a}}$ and $B_{x}$ have the
proper signs, this sweep results in an enhancement of the
non-uniformity. Figure 5a shows the effect of such a
back-and-forth sweep on the magnetic induction $B_{x}$ measured by
a sensor located close to the center of the sample. Two curves are
presented: a reference curve in which $H_{a}$ was swept between
$-6$T and 6T at a constant sweep rate; and the 'back-and-forth
sweep' (BF sweep) curve in which $H_{a}$ was swept up from $-6$T
to 3.9T and then was swept back and forth between 3.9T and 3.69T.
The reference curve shows a dip at 3.9T, signifying resonant
relaxation of the magnetization. The range of the back-and-forth
sweep was chosen to cover only half of this dip, at which both
$\frac{\partial B_{x}}{\partial H_{a}}$ and $B_{x}$ were negative.
During the back-and-forth sweep the magnetic induction $|B_{x}|$
at $H_{a}$ = 3.9T increased to a value 3.5 times larger than its
corresponding value at the reference curve. The strong increase in
$|B_{x}|$ above the saturation value signifies a strong
enhancement of the non-uniformity inside the sample, since $B_{x}$
is proportional to the local spatial derivative of the
magnetization, $\frac{\partial M}{\partial z}$, in the vicinity of
the sensor.
\begin{figure}
\includegraphics[width=0.42\textwidth]{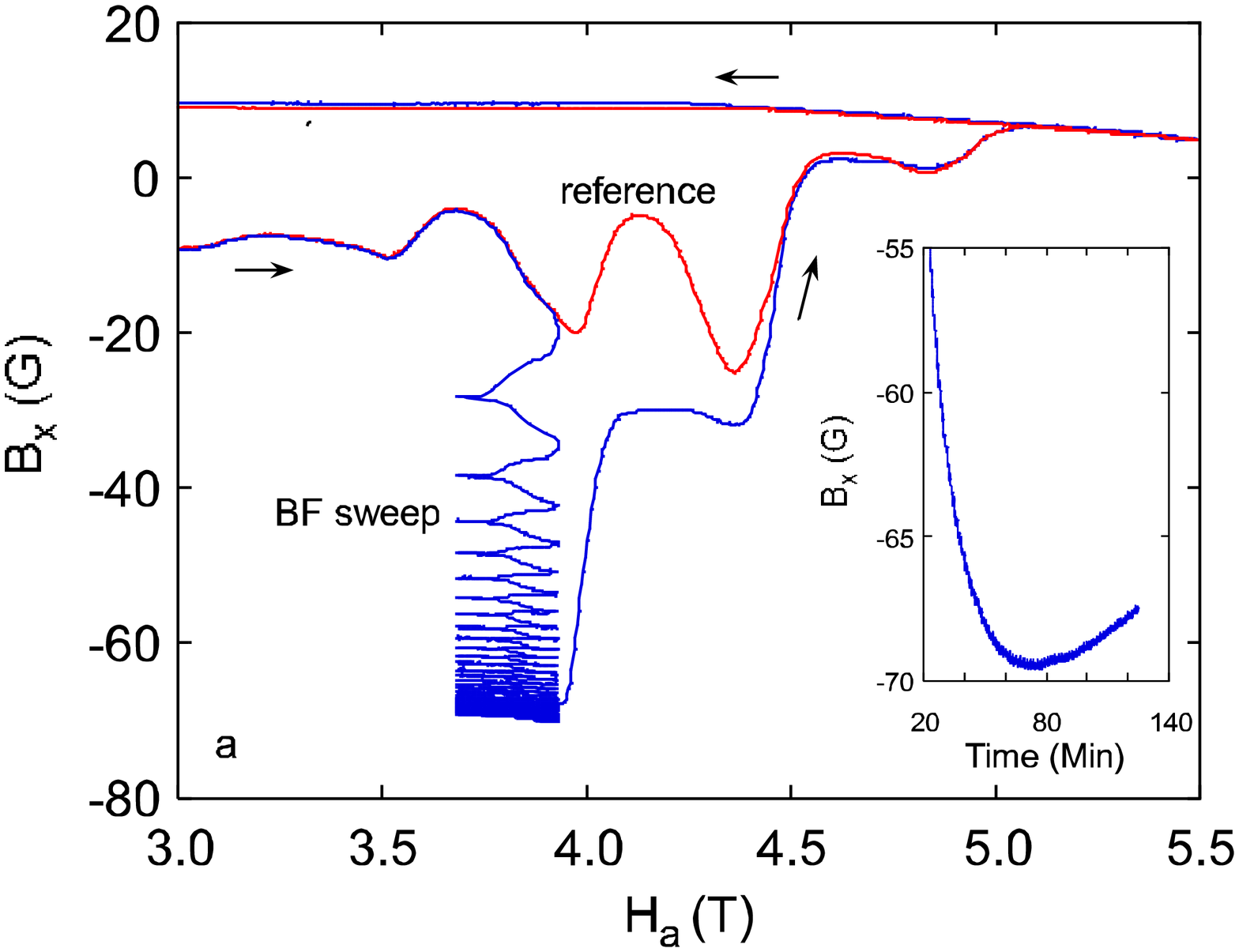}\\
\includegraphics[width=0.425\textwidth]{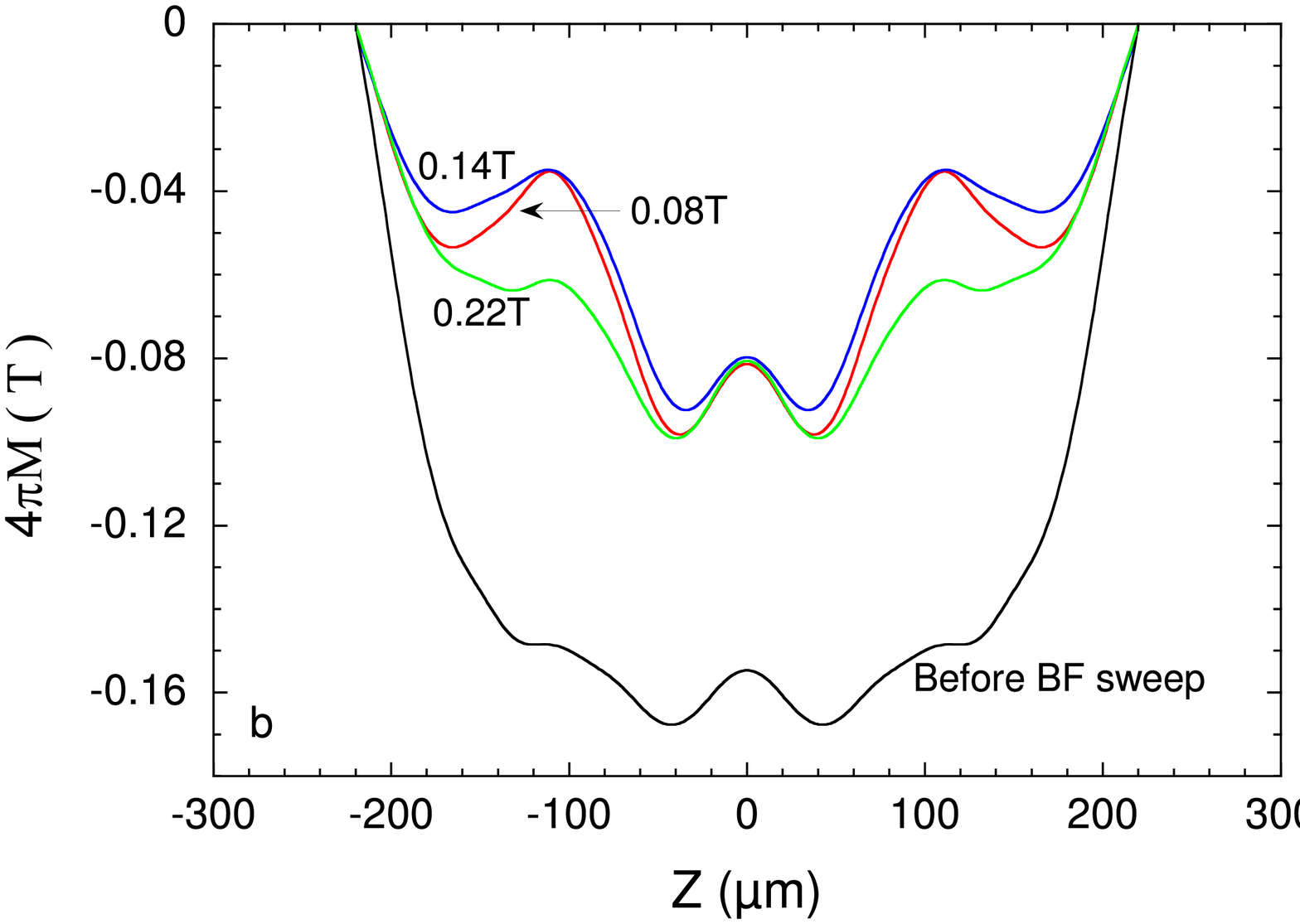}
\caption{\label{fig5} Local hysteresis loops $B_{x}(H_{a})$,
measured close to the center of the sample. During the back and
forth sweep the magnetic field  $B_{x}$ at $H_{a} = 3.9$T
decreases to $-70$G, which is 3.5 times larger than its
corresponding value at the reference curve ($-20$G). The inset
shows the evolution of $B_{x}$ as a function of time during the BF
sweep. (b) Magnetization profiles $M(z)$ showing the enhancement
of the non-uniformity by the BF sweep. The lowest curve shows the
magnetization at 3.68T before the BF sweep. The upper curves
correspond to three different measurements in which the BF sweep
was carried out with different amplitude of the BF cycle (0.08T,
0.14T, 0.22T)}
\end{figure}

Figure 5b demonstrates the enhancement of the non-uniformity in
the magnetization profile $M(z)$ by the BF sweep. The lowest curve
shows the magnetization at 3.68T just before sweeping the field
back and forth over part of the resonance. At this point some
non-uniformity has already developed. The three other curves
correspond to three different measurements in which BF sweeping
was carried out with three different amplitudes. In all
measurements the cycle starts at the same field of 3.7T, and each
measurement spans a different part of the transition. The most
significant enhancement of the non-uniformity is displayed for the
smallest amplitude of 0.08T, where the number of molecules that
relaxed during the BF sweep strongly differs between the center
and the edge of the sample. The effect is smaller for an amplitude
of 0.14T, and yet smaller for an amplitude of 0.25T. While
suggesting the possibility for magnetization non-uniformities that
grow with time, Eqs. (\ref{eq.5}) and (\ref{eq.6}) also indicate
that eventually, as relaxation progresses and the magnitude of
$M-M_{0}$ gets smaller, $\Gamma+4\pi(M-M_{0})\frac{\partial
\Gamma}{\partial B}$ must at some point become positive. Since
$\frac{\partial M}{\partial z}$ is approximately proportional to
the magnetic induction $B_{x}$ measured by our sensors, this
implies that when $H_{a}$ is swept back and forth for long enough,
the magnitude of $B_{x}$ must eventually decrease. This
expectation is indeed borne out in the measurement. The inset of
Fig. 5a presents the evolution of $B_{x}$ during the
back-and-forth sweeping as a function of time. The measured
$B_{x}$ initially increases in magnitude, but after about 70
minutes of sweeping, its magnitude starts to decrease. This change
happens when the difference between the relaxation rates on both
sides of the sensor changes sign. The source of this change in
relaxation rates may be the decrease of the magnitude of
$M-M_{0}$, or a change in the internal magnetic induction $B_{z}$
that changes both $\Gamma$ and $\frac{\partial \Gamma}{\partial
B}$. We are presently unable to determine the relative weight of
these two sources.

The saturation of the non-uniformity when $M - M_{0}$ becomes too
small is plausibly also the source of the effect of the BF sweep
on the following resonance. As shown in Fig. 5a, the BF sweeping
of the field in the range between 3.69T and 3.9T changes the dip
seen in the reference curve at 4.3T into a step. Within the
framework of Eq. (\ref{eq.5}), the non-monotonic behavior of
$B_{x}$ around 4.3T, shown in the reference curve, signifies an
enhancement of the non-uniformity followed by its suppression. In
contrast, the almost monotonic step observed in the BF curve
indicates that as a consequence of the back-and-forth sweep, the
resonance at 4.3T is traversed with
$\Gamma+4\pi(M-M_{0})\frac{\partial \Gamma}{\partial B}$ remaining
positive, such that no enhancement of the non-uniformity takes
place.

To summarize, we used a set of Hall sensors to measure the local
magnetic response of crystals of the molecular magnet
$\mathrm{Mn_{12}}$-acetate. Our measurements allowed us to determine the
local magnetization in different regions of the crystal and its evolution
as the crystal is driven through a hysteresis loop. We find significant
non-uniformities, which are larger when resonant tunnelling takes place,
but do not disappear when out of resonance. We explain how these
non-uniformities result from the dipolar interaction between molecules.
We show that the non-uniformity of the magnetization may be enhanced by
sweeping the externally applied field back and forth through a properly
chosen range. This method carries the potential for local manipulation of
the magnetization profile inside the sample.

The work at City College of New York was supported by NSF grant
DMR-0451605, and the work at WIS by the Israel Science Foundation
Center of Excellence (grant No. 8003/02) and by the US-Israel
Binational Science Foundation (BSF grants 2002238 and 2002242).

\end{document}